\documentclass[10pt, conference, compsocconf]{IEEEtran}

\usepackage{cite}

\ifCLASSINFOpdf
  \usepackage[pdftex]{graphicx}
\else
   \usepackage[dvips]{graphicx}
\fi
   
\usepackage{amsmath}
\usepackage{amsfonts}
\usepackage{algorithmic}
\usepackage[ruled,linesnumbered]{algorithm2e}
\usepackage{array}
\usepackage{fancyhdr}

\ifCLASSOPTIONcompsoc
  \usepackage[caption=false,font=normalsize,labelfont=sf,textfont=sf]{subfig}
\else
  \usepackage[caption=false,font=footnotesize]{subfig}
\fi

\usepackage{dblfloatfix}
\usepackage{url}
\usepackage{color}
\usepackage{caption}
\renewcommand{\tablename}{Tab.}

\begin{document}
%
\title{Ensuring Domain Consistency in an Adaptive Framework with Distributed Topology for Fluid Flow Simulations}

\author{\IEEEauthorblockN{Christoph Ertl, Ralf-Peter Mundani and Ernst Rank}
\IEEEauthorblockA{Chair for Computation in Engineering\\
Technical University of Munich, Germany\\
Corresponding author contact: christoph.ertl@tum.de}}

\maketitle

\begin{abstract}
Top-tier parallel computing clusters continue to accumulate more and more computational power with more and better CPUs and Networks. This allows, especially for environmental simulations, computations with larger domain sizes and better resolution. One of the challenges becoming increasingly important is the decomposition and distribution of the overall work load. State-of-the-art parallel codes usually use solutions that involve complete knowledge of the domain topology, which will lead to communication and memory bottlenecks when computing very large domains.
To meet this challenge, the authors propose a new strategy for decentralised domain management, based on a proven hierarchic data structure. On the way of developing a framework where individual sub-domains only have local knowledge of their surroundings, this contribution describes the communication patterns used in ensuring a consistent domain, without the need for expensive global broadcast messages. Furthermore, the routines necessary to deal with adaptive changes in domain topology, due to refinement, coarsening and migration of sub-domains to different computational resources, are discussed in detail.
\end{abstract}

\pagestyle{empty}
\thispagestyle{fancy}
\lhead{}
\chead{}
\rhead{}
\lfoot{This work has been submitted to the IEEE for possible publication. Copyright may be transferred without notice, after which this version may no longer be accessible.}
\cfoot{}
\rfoot{}

\flushbottom

\section{Introduction}
With the continued evolution of massive parallel systems, the size of the solvable problems grows accordingly. However, when computing on very large computational domains also new challenges surface, one of which is the efficient distribution of sub-domains to cores. One key part of parallel numeric codes especially in the context of adaptive mesh refinement is the domain management. This includes the subdivision of the domain, its distribution  and the continued balancing of workload to the computational resources during runtime. All this while keeping relations in check that are important for a fast and efficient computation, like neighbourhood relations and the number of data transfers between the individual domains. In adaptive settings, in which the topology of the domain is subject to change, the distribution of subdomains and therefore computational cost, while still maintaining these relations must be updated regularly.

State-of-the-art codes use space filling curves (SFC) to compute a linearisation of the individual subdomains. When cut into balanced partitions, the distributions gained from SFCs are proven to exhibit desired characteristics in terms of keeping neighbouring domains on physical close cores when distributed in a parallel network, ensuring fast message transfer \cite{bader2012space}. In addition, SFCs can also be computed in parallel, making their use a prime option in that application area \cite{luitjens2007parallel}. 

For large problems with complex domain topologies though, the memory requirement of the domain topology itself may hinder the actual computational work done on the cores, up to the point where it cannot be stored in its entirety any more. One remedy is the use of one of the cores or a group of processes be solely responsible for the bookkeeping of the domain topology, including taking care of all management tasks such as the ordering of refinements, coarsenings or migration of grids to a different core in case of a load imbalance. This strategy has been applied successfully by the works of J\'er\^ome Frisch \cite{bib:Frisch2014}. However, extensive testing has shown possible bottle-necks of this approach. The memory requirement for domain topologies of massive simulation tasks, for example when modelling the flooding of a city, are even too large for a dedicated domain management entity. Additionally, since all cores responsible for actually computing results must communicate regularly with the manager, communication channels will be overburdened. 

As remedy, the authors propose a newly developed strategy for a decentralised domain management and load-balancing based on the proven concepts of the aforementioned work. Key part of this concept is a heuristic that determines the best targets for the transfer of sub-domains on cores, weighting an optimised balance in terms of computational work with minimising communication cost when applying stencil based computational kernels. This paper describes the ongoing process of developing and implementing the communication routines that are necessary to achieve a decentralisation without global knowledge of the domain but only local domain view. Including the initial generation and distribution and the routines that are necessary to update the concerned subdomains when alterations are performed.

The remainder of this paper is structured as follows. The next section introduces the foundations kept from the work of J\'er\^ome Frisch. Particularly a short outline of the main application area of fluid dynamics with its governing equations and discretisation and the hierarchic data structure based on space-trees, especially designed for use in highly parallel environments. The third section presents the main contribution of the present article. First of all the facilities for generating and distributing the domain initially are discussed, followed by a closer look on the communication patterns used in ensuring the consistency of the domain after local refinement, coarsening and the migration of subdomains of the simulation space. How the local update routines are conceived for the three aforementioned domain modifications will be highlighted afterwards. To verify and compare the new update routines with the former approach, time measurements were conducted on a medium-tier cluster. The test setup and the results are illustrated in the second last section. The last section closes with a short summary and a broader view of the present contribution within the scope of the planned prospective work.

\section{Foundations}
The framework at hand is conceived as a derivative of an earlier framework developed at the Chair of Computation in Engineering. As such, the application area and its key characteristics are kept and where necessary enhanced. This includes specifically the data structure, and the solver concept. Nevertheless, the new framework was written from scratch and the former code base was completely revised, improving algorithmic design where deemed applicable. 
The present section will introduce the kept concepts in more detail and gives the interested reader the related citations for a comprehensive reference. The framework was designed for solving large scale fluid flow problems on massive parallel architectures. Subsequently the data structure and its particular applicability for distributed applications is illuminated. 

\subsection{Data Structure}
\begin{figure}[t]
\centering
\includegraphics[width=0.2\textwidth, height=3.5in]{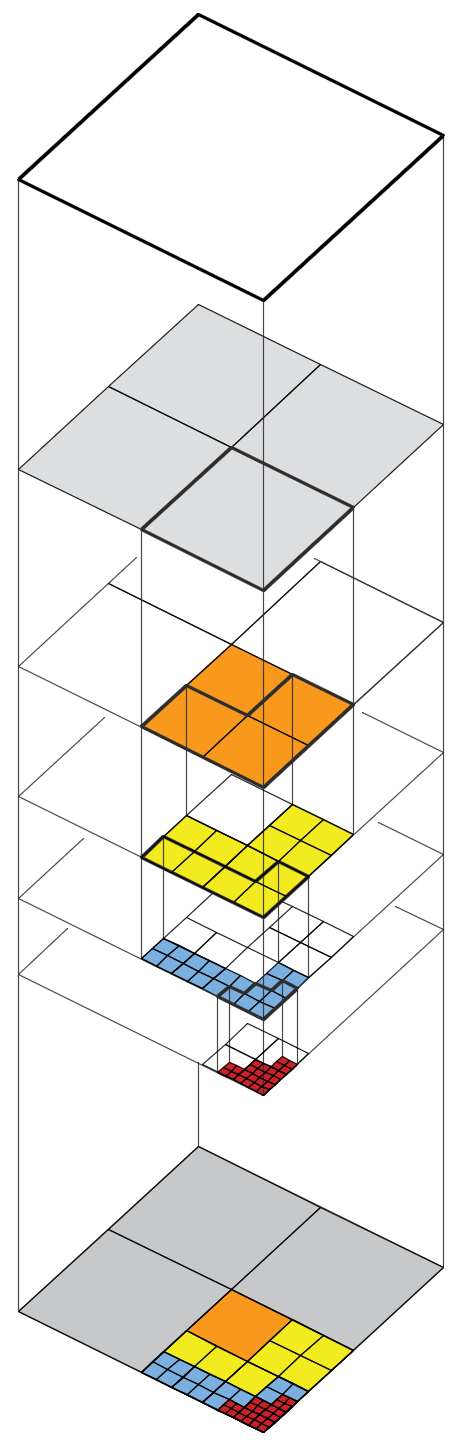}
  \caption{Example 2D data structure, adaptively refined up to depth 5}
  \label{fig:structure}
\end{figure}

To be able to be distributed, the domain is divided into 3D block-structured regular Cartesian grids. Each of the grids regardless of their physical extent encloses the same amount of cells which constitute the entities at which the entries of the solution fields (velocities, pressure, temperature etc.) are stored. The grids are generated following the general idea of space-trees (with quadtrees as 2D and octrees as 3D representatives, see also \cite{bib:babufrmu2002}) starting from a single root grid, representing the complete simulation space on depth $0$ of the tree. This root grid is then subdivided by $r_x \times r_y \times r_z$ spawning grids on depth $1$. By subsequently subdividing grids on the deepest, that is the one with the largest depth, the domain is represented by more and physically smaller grids, representing the domain in an increasingly finer resolution, until a predefined depth $d_{max}$ has been reached. Fig. \ref{fig:structure} illustrates an example data structure, for simplicity in 2D. The grids are successively refined using a bisection in the two cardinal directions up to a depth of $5$. Additionally, the data structure supports also adaptive configurations, enabling to represent regions of interest, or where numerically necessary with a finer resolution. 

\begin{figure}[t]
\centering
\includegraphics[width=0.4\textwidth]{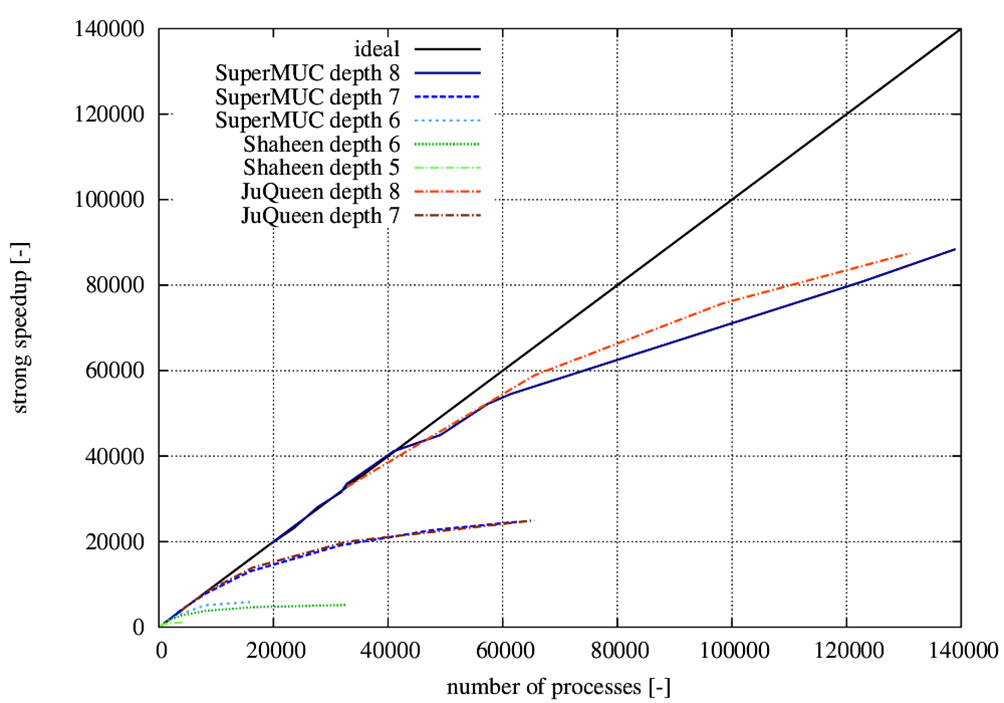}
    \caption{Strong speed-up values plotted against the number of processes}
    \label{fig:speedup}
\end{figure}

\begin{figure}[t]
\centering
\includegraphics[width=0.4\textwidth]{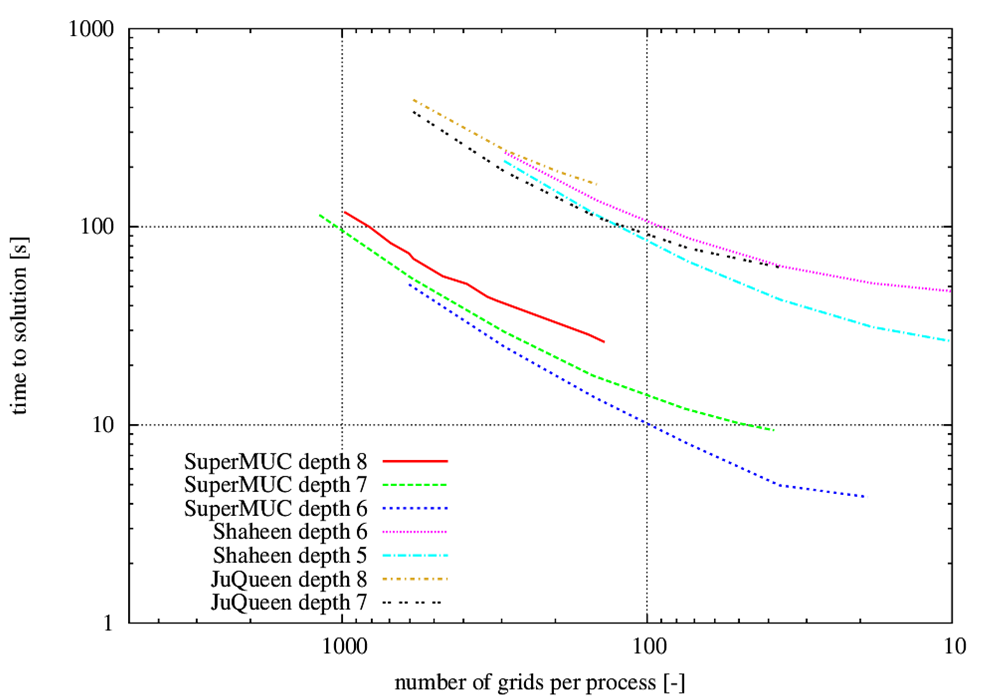}
    \caption{Time-to-solution (one full time step) plotted against the number of grids per processes}
    \label{fig:timetosolution}
\end{figure}

The unique characteristic of the structure is that coarser representations of the domain are not discarded. Custom tailored to this, a parallel multigrid-like solver according to Brandt \cite{bib:brandt1977} was developed. The solver benefits from not having to compute coarser grid representations of the solution field as they are readily available. Additionally, it can reuse the communication facilities used in refreshing the data values across the hierarchic levels, making it readily available for parallel use. In Fig.~\ref{fig:speedup} and Fig.~\ref{fig:timetosolution} the strong speed-up as well as the time-to-solution of solving a Poisson equation for different resolutions up to $256 \times 256 \times 256$ (depth 8) grids on the finest level with approx.\ 707 billion unknowns on different HPC systems is shown. Details about the solver and all performed analyses and comparisons can be found in \cite{bib:frmura2013}.

In addition, the keeping of the coarser hierarchy levels allows for an elegant way of visualising simulation results, while limiting the amount of data in need to be processed. Main idea is to only select subsets of data based on a chosen region of interest while varying the resolution of displayed data points equivalent to the size of the view. A more thorough look on this technique  can be found in \cite{bib:mufrvara2015}.

\subsection{Message Passing and Grid Identification}
\label{subsec:UID}
For parallel functionality, the framework relies on the Message Passing Interface (MPI) in version 3.1 allowing communication between processes in a distributed memory environment \cite{bib:MPI}. Within the distributed environment each participating process is identified by its rank, a 32 bit unsigned integer. In the following, processes and ranks are used interchangeably. Usually one process is spawned per core. Because each core has its own memory, when information from a remote process is needed, explicit transfer of data via messages is necessary. 

To uniquely identify a grid in the domain each grid is assigned a unique Identifier (UID). This UID is a long integer consisting of 64 bits, whereas the first 32 bits from left to right encode the rank on which the grid resides. The next 23 bits are used to encode the grid identifier (GID), a number exclusive for a grid on a certain rank. Finally, the last nine bits are used to encode the position of the grid in its super-grid, with always three bits signifying one of the cardinal directions. The combination of those nine bits is also called a grid's hash. Fig. \ref{tab:UID} illustrates this composition of the UID. The combination of GID and the hash is also called a tag. 

\begin{figure}
	\centering
	\begin{tabular}{c | c | c | c | c }
		rank & \multicolumn{4}{c}{tag}\\
		32 & \multicolumn{4}{c}{32}\\
		rank & GID & \multicolumn{3}{c}{hash}\\
		32 & 21 & \multicolumn{3}{c}{9}\\
		rank & GID & $p_{k}$ & $p_{j}$ & $p_{i}$ \\
		32 & 21 & 3 & 3 & 3\\
	\end{tabular}
	\caption{Construction of the UID and naming of individual construction elements}
	\label{tab:UID}
\end{figure}

This concludes the foundations built upon. The following chapter will discuss current work towards the goal of self-managing subdomains on processes without the need for a centralised management facility.

\section{Domain Management}
To develop a framework where the executing processes are able to successfully manage their own share of workload, including the redistribution of such when necessary, without the global knowledge of domain topology, two main challenges have been identified. The first of which is a suitable communication scheme used in keeping the domain consistent across all subdomains. That means every grid knows its neighbours both in a geometrical and a hierarchical sense, which is needed for communicating bordering values, residuals and errors used in the stencil computations and multigrid restriction and prolongation operations. This neighbouring information has to be updated regularly to keep track of changes in the domain topology due to refinements and deletion caused by numerical considerations or due to migrations to balance computational workload.  

The second challenge is the development of a load-balancing strategy, tailored to the presented data structure and limited domain knowledge. Key part of this concept is a heuristic that determines the best targets for the transfer of sub-domains on cores modelled after a diffusion process. Diffusion based load-balancing schemes in the context of distributed processing can be found at \cite{bib:cybenko89}. This heuristic is part of current research and, thus, will not be elaborated in detail here.

Before introducing the main contribution of this publication, the communication routines necessary for a decentral domain management, the subsequent section briefly highlights the shortcomings of the formerly used centralised approach utilising one of the processes as a dedicated topology repository and manager. Furthermore, the reasoning behind the characteristics of the new approach is highlighted. 

\subsection{Lessons Learned and Design Goals}
As already mentioned, large-scale testing on SuperMUC and JuQueen has shown bottlenecks of the previously used centralised domain management facilities. On SuperMUC the pruned tree network aggravated the regular communication of every computing process with the management instance. One remedy was to introduce duplicate management instances, each responsible for a reduced number of processes, effectively reducing the communication load per instance. However, the price to pay is an increased effort in synchronising each instance of the domain manager as well as the need to pin the managers responsible for a cluster of processes onto the same rack as those processes -- the later not in all cases possible. 
On the JuQueen, competition of network links is no problem due to the very efficient 5D torus network. However the comparably small memory per core does not allow to store the complete topology on one process, making the use of the domain manager impossible. The only remedy to run a simulation on the machine was to compute the distribution of the topology on a different machine, then having each process read in a configuration file. This results in a fixed topology where no adaptive load-balancing and subsequent topology changes are possible. 
The only viable approach to solve the afore-mentioned problems is a system without the central domain manager. This approaches is fairly common among numerical frameworks, having the completely domain topology be available on all participating processes has some advantages and drawbacks also. The complete knowledges allows the computation of an optimal global load-balance, with the most common approach being the use of a space-filling curve distribution. A large overhead of communication is necessary however, to keep the domain information concurrent on all processes, as well as the problem with limited memory for storing the information is present. In this case even more prevalent, since memory occupied by topology information limits the available space for actual simulation data and decreases computational efficiency. These shortcomings were addressed by the development of the central domain manager, which postponed them, but was not able to solve the issues conclusively. 
The authors therefore propose a system, without a central manager, where individual processes only have knowledge of their immediate surroundings. Advantages of this approach are a fixed memory overhead of the domain information per computational grid, as well as an upper bound of necessary communication per grid. Communication needs only be established between processes that hold neighbouring grids, broadcast operations are completely avoided. A grid in 3D has at most six geometrical neighbours, one hierarchical parent plus a fixed number of hierarchical children. In the worst case all neighbours of a grid lay on different processors, requiring network communication. Therefore, an almost infinite scalability in a weak sense, i.e. when increasing the problem size while keeping the number of grids per process constant, is expected.

\subsection{Domain Generation and Initial Distribution}
Even though for the load-balancing as it will be conceived, a complete domain view is not necessary, the initial distribution is carried out from a central location, that is from the master process or rank zero. The domain is created by generating a single root grid, representing the complete simulation domain. The root grid will be successively refined in the three cardinal directions up to a defined maximum depth. This may be carried out uniformly, such that the entire domain exhibits the same level of detail, or adaptively, meaning only parts of the domain which are of special interest or need to have a finer resolution due to numerical accuracy are resolved further. The grids generated in this step are hulls, containing only topological information, that is information about their neighbours, their location and physical extent. The memory for the simulation data, for example velocities, pressure and temperature, will not be allocated before the distribution to their respective ranks has been carried out.

After the initial creation, the master rank calculates a sorting according to a space-filling curve using the Morton or depth first scheme. The linearised order is then cut into equal amounts to be distributed among all participating processes. Before the grids are actually distributed, the master process updates the neighbourhood information with the new UIDs of all grids. This avoids an expensive additional back and forth communication, querying for the UID and is possible because the destination rank as well as the tag of a grid are known. Since no grids reside on remote ranks at this point, GIDs may be distributed simply in ascending order. 

Finally, the grids are migrated including their meta data to their respective destination ranks as determined by the Morton ordering. In this step, every participating rank must at least obtain one grid. This is necessary since communication is only established between ranks which share a neighbouring, sub- or super-grid. Load-balancing therefore cannot consider empty ranks which would lead to a waste of resources. 

This procedure also marks a deviation from the preceding framework. Due to geometry being read in at the master process, the memory for every grid's cell type i.e. fluid, solid etc. had to be allocated on the master process, further limiting the problem size. Now geometry is read by every process where cell types are allocated after distributing the grid hulls. The possible load imbalance stemming from grids having uneven load is compensated afterwards using the diffusion load-balancing.

\subsection{Communication Pattern}
At this stage in development, it was deemed to be more advantageous to enforce a strict separation of communication and computation phases. For once, this ensures consistency and reliable results when developing the decentral communication. Furthermore, the separation of the computational kernel from the underlying communication, also allows experts in numerical fluid simulations but not in high-performance computing to implement their computational kernels, without having to be knowledgeable in both regimes, making the framework easily extensible. 

The basic premise of the decentral scheme is to completely avoid expensive broadcast operations to distribute global information, as well as abandon a single or a few central repositories, that would have to be contended by the majority of processes trying to update their neighbourhood information. The most important paradigm in the scheme is that ranks need only to communicate with remote ranks with which they have neighbouring links to. Meaning that grids that lay on the origin rank are neighbours to grids on the remote rank, both in terms of hierarchy via super- or subgrids, and in terms of geometry as bordering grids on the same tree depth. In its first iteration the communication pattern was ordered in a regular fashion. Each rank owns an ordered vector \textit{RemoteRank}, consisting of all ranks with which it needs to communicate. 

\begin{algorithm}
	\caption{Structured communication pattern}
	\label{alg1}	
	\SetAlgoLined
		\Begin{
		\For{ $remoteRankIter < rank$ }{
			receive update queries from remote rank \\
			send update queries to remote rank \\							
		} 
		update all local neighbours \\
		$remoteRankIter$++ \\	
		\For{ $remoteRankIter\ != size $}{
			send update queries to remote rank \\	
			receive update queries from remote rank \\
		} 	
		}
\end{algorithm}

In algorithm \ref{alg1}, the scheme is outlined. An Iterator \textit{RemoteRankIter} is used in traversing the vector in a linear way. First, all ranks post a blocking receive followed by a blocking send for all ranks that are smaller than their own rank integer. Since domain updates between grids might as well happen for grids on the same rank, those are dealt with next. To complete the pattern, a blocking send followed by a receive is posted for all ranks larger than one's own, in reverse order as before. 

Tab. \ref{tab:6ranks} visually illustrates the resulting pattern when six participating ranks, where every rank needs to communicate with each other. The columns signify individual ranks, the rows the communication operations. Entries in the table signify the corresponding rank integer with which communication happens. The color coding serves to increase the visibility of communicating ranks. One can immediately observe the ordered pattern with no danger of a deadlock. 

\begin{table}[h]
\centering
\begin{tabular}{c c c c c c c}
stage & \multicolumn{6}{c}{rank} \\
 & 0 & 1 & 2 & 3 & 4 & 5 \\
\hline
1 & {\color{blue} 1} & {\color{blue} 0} & - & - & - & - \\
2 & {\color{blue} 2} & - & {\color{blue} 0} & - & - & - \\
3 & {\color{blue} 3} & {\color{red} 2} & {\color{red} 1} & {\color{blue} 0} & - & - \\
4 & {\color{blue} 4} & {\color{red} 3} & - & {\color{red} 1} & {\color{blue} 0} & -\\
5 & {\color{blue} 5} & {\color{red} 4} & {\color{green} 3} & {\color{green} 2} & {\color{red} 1} & {\color{blue} 0} \\
6 & - & {\color{red} 5} & {\color{green} 4} & - & {\color{green} 2} & {\color{red} 1} \\
7 & - & - & {\color{green} 5} & {\color{magenta} 4} & {\color{magenta} 3} & {\color{green} 2} \\
8 & - & - & - & {\color{magenta} 5} & - & {\color{magenta} 3} \\
9 & - & - & - & - & {\color{cyan} 5} & {\color{cyan} 4} \\
\end{tabular}
\caption{Regular communication pattern with six ranks where every rank needs to exchange data with each other rank}
\label{tab:6ranks}
\end{table}

\vspace{-7pt}

However, it also is evident that this pattern is not optimal due to the idling of certain processes in specific communication stages. One quick way of improving the efficiency is to join suitable communication stages. This is illustrated in Tab. \ref{tab:6ranksOpt} for six participating ranks.

\begin{table}
\centering
\begin{tabular}{c c c c c c c}
stage & \multicolumn{6}{c}{rank} \\
 & 0 & 1 & 2 & 3 & 4 & 5 \\
\hline
1 \& 7 & {\color{blue} 1} & {\color{blue} 0} & {\color{green} 5} & {\color{magenta} 4} & {\color{magenta} 3} & {\color{green} 2} \\
2 \& 8 & {\color{blue} 2} & - & {\color{blue} 0} & {\color{magenta} 5} & - & {\color{magenta} 3} \\
3 \& 9 & {\color{blue} 3} & {\color{red} 2} & {\color{red} 1} & {\color{blue} 0} & {\color{cyan} 5} & {\color{cyan} 4} \\
4 & {\color{blue} 4} & {\color{red} 3} & - & {\color{red} 1} & {\color{blue} 0} & -\\
5 & {\color{blue} 5} & {\color{red} 4} & {\color{green} 3} & {\color{green} 2} & {\color{red} 1} & {\color{blue} 0} \\
6 & - & {\color{red} 5} & {\color{green} 4} & - & {\color{green} 2} & {\color{red} 1} \\
\end{tabular}
\caption{Optimised communication pattern with six ranks where every rank needs to exchange data with each other rank}
\label{tab:6ranksOpt}
\end{table}

Furthermore, it can be shown that for even counts of participating processes a complete reordering yields a perfect solution. For odd numbers an almost perfect solution, with one process idling during each stage is achievable. This is accompanied by a drastic increase in algorithmic complexity however. For a code framework that is consistently requiring communication from every rank with all others, the effort in developing and implementing such a pattern would certainly be warranted. However, for the framework at hand, ranks only need to communicate with ranks that hold specific grids, severely limiting the number of exchange pairs. In addition, that relationship is subject to frequent changes due to load-balancing and the resulting migration of grids. The computation of an optimal pattern requires global knowledge of all pairs, rendering it unobtainable in the present framework.

For the problem at hand there are two reasonable remedies. The first would be to rely on buffered non-blocking communication, which from an implementation point of view is readily available since the same pattern as discussed earlier can be used. The advantage is that the strict ordering is abolished, however for the cost that every communication has to be supported by a large enough buffer. The buffer memory limiting the amount of grids per rank leads to a decreased computational efficiency because there is less data available to process per rank. The second possible remedy is using MPI's one-sided communication routines in which the origin process is assigned a window in the remote rank's memory. It therefore can directly interact with the remote machines' memory without handshaking, leading in theory to a much more efficient communication pattern. Here, special care to ensure concurrency due to clashing operations has to be taken care of. The later certainly increases algorithmic complexity, however promises the best results in communication efficiency and memory requirement. As such, it is planned to be implemented and tested in the next overhaul of the communication routines.

This concludes the section describing the pattern for exchanging messages decentrally in the domain. The subsequent section illustrates the actual update routines, responsible for exchanging the possible domain alterations, namely the refinement of grids, their deletion as well as their migration to a different rank.  

\subsection{Domain Update Algorithms}
The three possible alterations that may change the topology of the domain are refinement, deletion and migration of a grid to a different process. The first two may stem from a user interaction, or from numerical reasons when more or less accuracy is needed. While grids are generated or deleted on a process, the computational workload per process changes. If it reaches a certain threshold, a load-balancing becomes necessary. Since no complete domain view is available, the balancing is realised using a diffusion-like heuristics, weighting an optimised balance in terms of computational work with minimising communication cost. Subsequently, grids have to be migrated to establish the computed balance.

Each process employs a so-called \textit{queryVector} for every process including itself, a list of 64 bit integers closely related to the aforementioned UID. These integers code the tasks, that are exchanged with the corresponding processes. In Fig. \ref{tab:query} the construction of a query is illustrated. The first 27 bits from left to right are unused at the moment. Two bits encode the three possible tasks. While the queryVector here does not include migration queries, the structure of the query is kept the same. The next three bits code the direction in which the neighbour that issues the query lies. These can range from zero to five for the six neighbours on the same hierarchic level, that is east, west, north, south, top and bottom. Two more integers six and seven signify a sub- or a supergrid. Via the following GID, a query is able to uniquely identify a grid on the remote rank which is affected by a change on the origin rank. Finally, if the subgrid of a grid is needed, the hash is used to identify it via its position in its local coordinate system.  

\begin{figure}
	\centering
	\begin{tabular}{c | c | c | c | c }
		unused & task & direction & GID & hash \\
		27 & 2 & 3 & 21 & 9\\
	\end{tabular}
	\caption{Construction of a query}
	\label{tab:query}
\end{figure}

\subsubsection{Refinement and Deletion}
Refinements naturally only happen at non-refined grids to increase the resolution of the represented domain. For deletion, removing of a non-leaf grid would make sense, leads however to a cascade of deletions of all descending grids. When a rank is affected multiple times during such a deletion cascade, one process can only identify all affected grids by checking the boundary box of the highest hierarchical grids and intersecting it with all other grids residing on it, leading to a variable overhead. Furthermore, if this cascade spreads to processes further down the hierarchy with already finished communication during the current cycle, one full cycle is not sufficient to carry out all necessary operations. Restricting deletion to only leaf grids, allows both update operations to be carried out within one full communication cycle without additional overhead. Furthermore, refinement and deletion neither cause the communication pattern to change nor influence each other. All refinement and deletion queries are grouped in one \textit{queryVector} an are carried out within one cycle where the ordering of operations is arbitrary and conceived in order issue. 

If a grid is refined, the task of the update function is to determine the neighbours of the newly generated subgrids on the same tree level. Possible neighbours are the subgrids of neighbouring grids of the grid that is refined. Therefore, a refinement query asks remote ranks if neighbours of the grids are refined and if applicable, send back the respective neighbouring subgrids. The set of queries spawning from the refinement of a grid therefore includes the GIDs of its geometric neighbours. The task is coded with two bits according to Fig. \ref{tab:query}. Since refinements only need to consider neighbours on the same tree-depth, directions can range from zero to five and the hash is unused.

\begin{algorithm}
	\caption{Routine for sending refinement and deletion queries}
	\label{alg2}
	\SetAlgoLined
	\Begin{
		send the \textit{queryVector} \\	
		receive all new neighbours \\
		update neighbour information \\
		send all corresponding neighbours with positive queries\\						
		}
\end{algorithm}

\begin{algorithm}
	\caption{Routine for receiving refinement and deletion queries }
	\label{alg3}	
	\Begin{		
		receive the \textit{queryVector} \\	
		\ForEach {query in the \textit{queryVector}}{
			\If {task is deletion}{
				find grid in question \\
				delete neighbour reference \\
			}
			\ElseIf{task is refinement}{
				\If{grid is refined}{
					put neighbouring subgrids in a vector for collective transmission \\
			}
			}
			}
		send all found neighbours \\
		receive all new neighbours \\
		update neighbour information \\	
		delete duplicate queries \\				
		} 
\end{algorithm}

If a grid is deleted, similarly, the task of the update function is to determine the grids that are affected by the deletion and inform them to delete their neighbour reference. Again task and GID of the affected grids are coded according to Fig. \ref{tab:query}. Directions can range from zero to six to signify the deletion of a geometrical neighbour or the subgrid of a grid. The later case additionally makes use of the hash to identify the correct subgrid. 

Algorithms \ref{alg2} and \ref{alg3} illustrate the routines carried out for updating the neighbourhood information of all concerned ranks due to refinements and deletions. See also the communication pattern \ref{alg1} for the ordering of the sending and receiving routines. Algorithm \ref{alg2} depicts the view from the origin process. It first sends all its refinement and deletion queries to the corresponding remote rank. Afterwards it receives all positive answers from the refinement queries, including the UIDs of the subgrids that qualified as new neighbours and updates its own grids accordingly. Since every grid that got a positive response on the origin rank has gotten a new neighbour, the converse is also true, meaning that the hits on the remote rank get new neighbours. These are collected and send to the remote rank in the final step of the algorithm.

On the remote side, depicted in algorithm \ref{alg3}, the corresponding operation to receive the \textit{queryVector} is posted. Afterwards, a loop traverses every query in the vector, classifying the task to be performed. For deletion, the affected grid needs to be identified and the neighbourhood reference in the direction given is erased. For refinement, a positive response is justified if the grid in question is refined. If true, the subgrids in the opposite direction as the query are collected in a so-called \textit{neighbourVector}. After all queries are processed, the collected neighbour UIDs are sent to the origin rank, followed by a receive of the previously mentioned converse relation. Using this information, the neighbouring relations are updated on the remote rank accordingly. Finally, in case a refinement is carried out on both sides of a neighbour pair residing on different ranks, the queries would be processed twice since both refinements cause queries to the respective remote rank. To avoid unnecessary work, the duplicate queries are erased from the \textit{queryVector} on the remote side.

\subsubsection{Migration}
For migration the picture looks different. Due to migration of grids, the relation between ranks is subject to change. Since ranks only establish communication to other ranks with which their grids share neighbourhood relations, the migration of grids leads to the need to establish communication with new ranks or abandon communication respectively. This means that migration cannot be handled within a single communication cycle but two are needed. In total three cycles are used to carry out one round of refinement, deletion and migration, with a load-balancing step before the migration. 

In one cycle, illustrated in algorithms \ref{alg4} and \ref{alg5}, the grids in question are actually migrated to the receiving rank, including all meta data and all simulation variables. The receiving rank then returns the newly generated UID for the grids. For every transfer, the origin rank keeps a list of all neighbours, that have to be informed of the change since the target rank is not yet able to inform those ranks by itself. Again the links are not yet established. The remote rank can only be selected out of the set that previously was in the communication rotation. The authors expect this to be beneficial in ensuring most of the communication, induced by the computational kernels when values at the grid borders have to be exchanged to be between physical close processes, comparable to a SFC distribution.

\vspace{-7pt}

\begin{algorithm}
	\caption{Routine for sending migration queries}
	\label{alg4}	
	\SetAlgoLined
		send all migration grids \\
		receive all new UIDs \\
\end{algorithm}

\vspace{-17pt}

\begin{algorithm}
	\caption{Routine for receiving migration queries}
	\label{alg5}	
	\SetAlgoLined
		receive all migration grids \\
		generate new UIDs \\
		send new UIDs \\	
\end{algorithm}

\vspace{-7pt}

In the second cycle, illustrated in algorithms \ref{alg6} and \ref{alg7}, the neighbourhood information on all ranks affected has to be updated. The main difficulty is posed by the fact, that multiple grids have been migrated from one origin, altering targets in the update list. It is therefore necessary to reference the update list for changes using the UID responses from the migration step. After the list has been revised, all targets can be informed of the migrations. This again makes use of the query structure, coding the task, the GID of the grid on the remote rank that needs to update its neighbourhood information, the direction and, if necessary, the hash to uniquely identify a subgrid. 

When all migrations and the resulting updates are completed, rendering all information in the domain consistent, all ranks update their list of ranks with which they share neighbouring pairs with.

\vspace{-7pt}

\begin{algorithm}
	\caption{Routine for sending update information}
	\label{alg6}	
	\SetAlgoLined
		update \textit{queryVector} with new UIDs \\
		send \textit{queryVector} \\
		send new UIDs \\
		update list of communication links \\
\end{algorithm}

\vspace{-17pt}

\begin{algorithm}
	\caption{Routine for receiving update queries}
	\label{alg7}	
	\SetAlgoLined
		receive all \textit{queryVector} \\
		receive new UIDs \\
		\ForEach {query in the \textit{queryVector}}{
			identify the grid in question \\
			update neighbour reference \\
			}
			update list of communication links \\
\end{algorithm}

\vspace{-7pt}

\section{Performance Measurements}
To be able to classify the new decentral communication routines, they were tested using a series of benchmarks against the former central approach. Tests were performed on an Intel Xeon E5-2697 v3 "Haswell"cCluster operated by the Leibniz Supercomputing Center of the Bavarian Academy of Sciences and Humanities. The cluster incorporates 384 nodes, with 28 cores per node and 10752 cores in total. The nodes are connected with Infiniband FDR14 interconnects. 

\begin{figure}[t]
\centering
\includegraphics[width=0.5\textwidth]{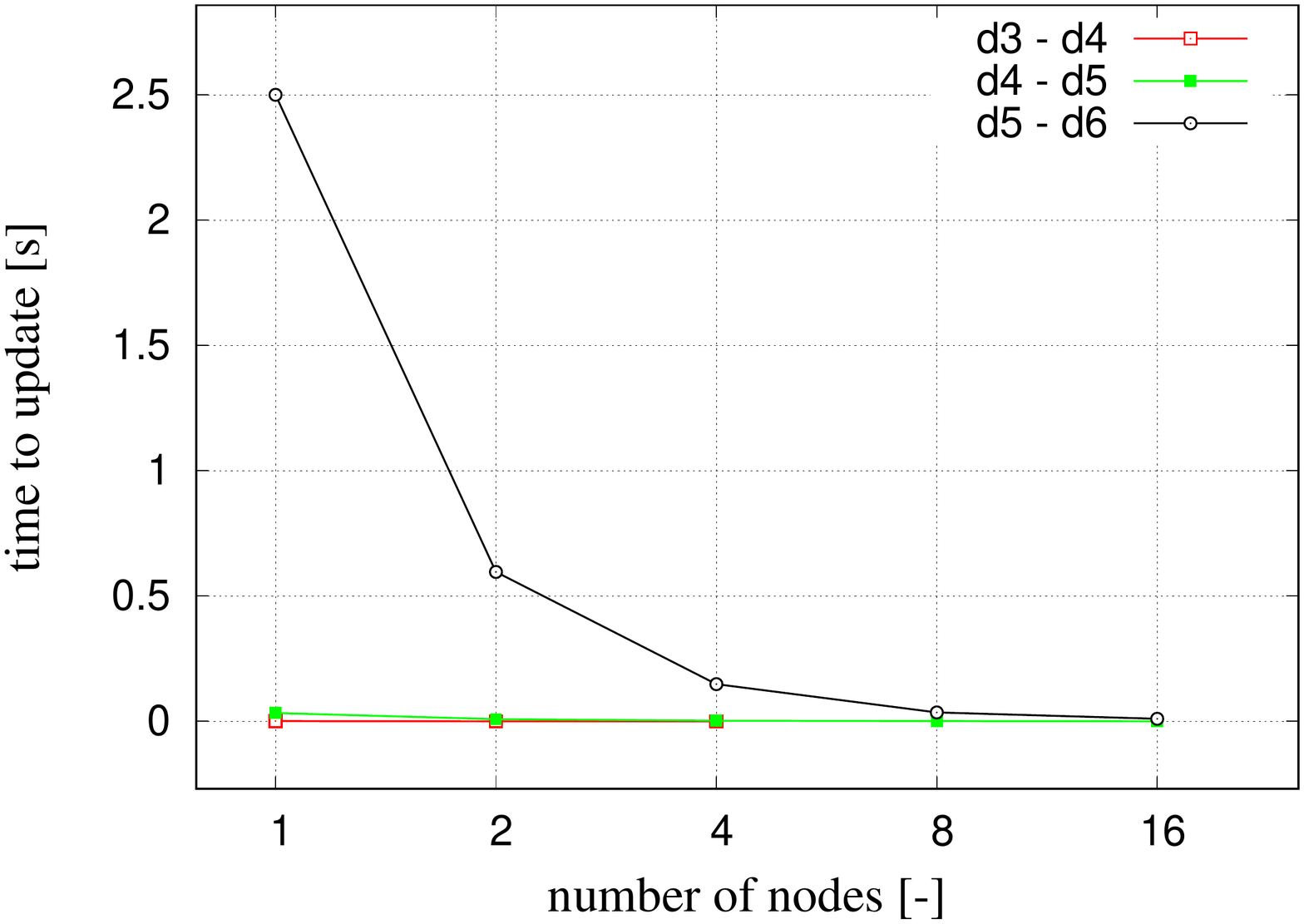}
    \caption{Update times plotted against the number of nodes for different depths when refining the complete domain, using the decentral update routines}
    \label{fig:s1}
\end{figure}

\begin{figure}[t]
\centering
\includegraphics[width=0.5\textwidth]{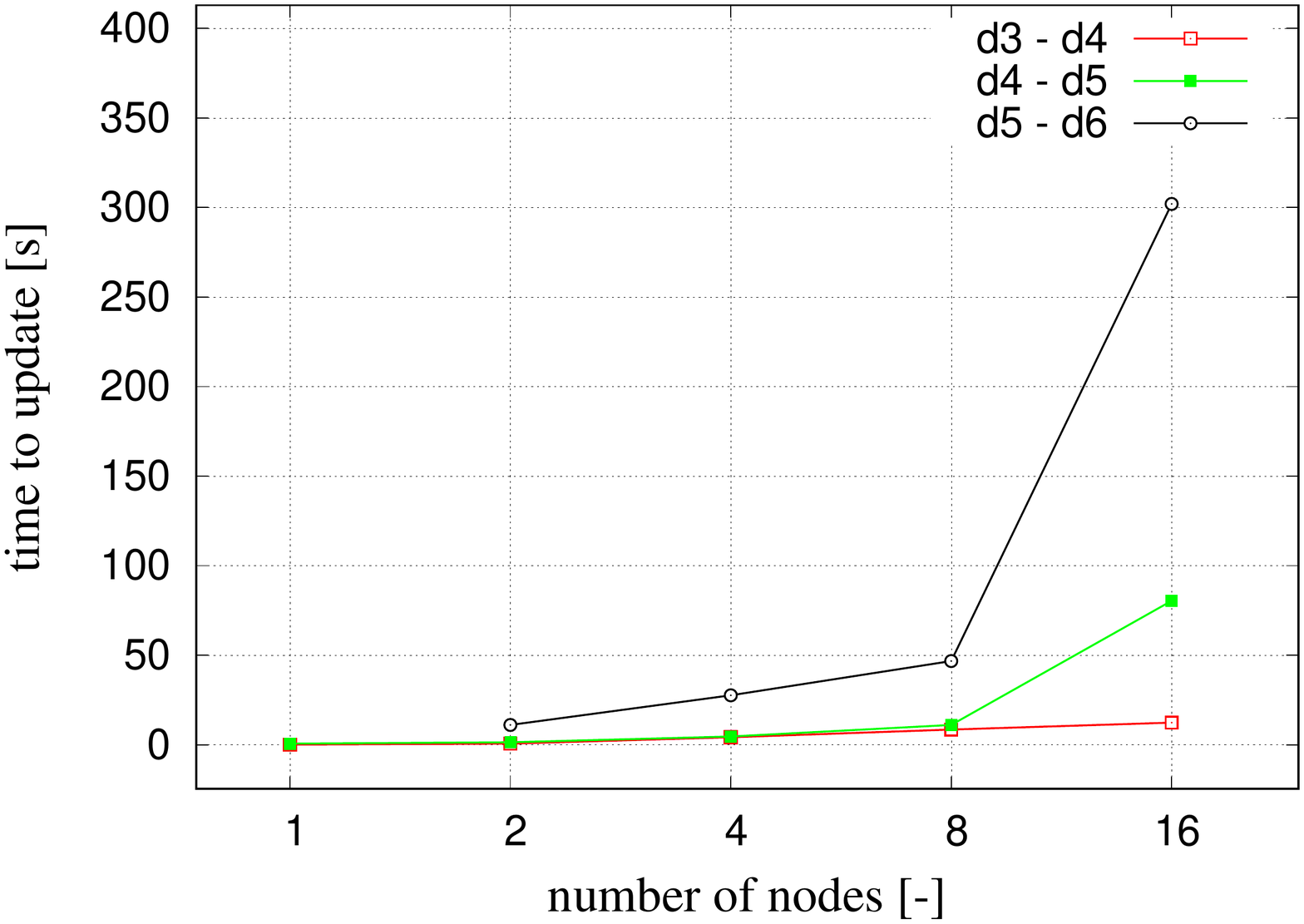}
    \caption{Update times plotted against the number of nodes for different depths when refining the complete domain, using the central update routines}
    \label{fig:s2}
\end{figure}

For the test setup a cubic domain was uniformly refined up to a prescribed depth for three test cases using a bisection in each of the cardinal directions. All grids were migrated to the participating processes using the morton space-filling curve distribution. The complete domain is then refined by one level, measuring the total exchange times due to communication caused by the refinement, i.e. querying and informing concerned neighbouring grids, updating their neighbourhood information accordingly. The chosen test cases include refinements from depth three to four, four to five and five to six. With 585, 4681, 37449 and 299593 grids in total for the respective levels. The measurements were then carried out on one, two, four, eight and sixteen nodes, or 28, 56, 112, 224 and 448 cores. The results are illustrated in Fig. \ref{fig:s1} for the decentral communication routines and Fig. \ref{fig:s2} for the former, central communication routines. 

It can clearly be observed that the decentral routines follow a decreasing trend with more nodes involved. With times decreasing from 2.5 seconds when using one node to around one millisecond when using 16 nodes and a refinement from depth five to depth six. This matches with the expectations that a grid introduces a bounded amount of communication to its respective process. When increasing the amount of nodes, respectively processes, the number of grids per process decreases, reducing the communication per process. In combination with the decentral communication, the overall update time is also reduced. In comparison, the central routines show the complete opposite trend. When refining the domain from depth four to five, the update time increases from 11 seconds to 300 seconds using 16 nodes. The fact that the number of grids per process is reduced when increasing the number of nodes is true as well. However, every process needs to query the neighbourhood server for changes concerning them, leading to the aforementioned communication bottleneck. 

In summary, the measurements show the expected excellent scaling properties of the decentral approach which were the motivation behind revising the domain management facilities. The last section concludes by summarising the main findings of this contribution and gives an outlook into future work.

\section{Conclusion and Future Work}
The present work described the current work on a framework for parallel fluid flow simulations based on a proven hierarchical data structure. This framework is designed to not require expensive broadcast operations or central domain management facilities, which limit the scalability when computing on current peta- or even future exascale machines. On the way the achieve this via self-managing subdomains, the authors have presented the communication patterns as well as the routines to ensure domain consistency when adaptive domain modifications become imperative due to numerical necessities. The facilities described here are subject to be continuously reworked. In case of the communication pattern, possible optimisations have been laid out.

Testing on a medium-tier machine on up to 448 cores has shown very promising results in terms of communication times caused by a refinement of a complete uniform domain of one level and supports the continued development in this direction. 

The next step is the conception of the load-balancing kernel, again restricted by only a local domain view of the individual subdomains, in conjunction with incorporating the unique characteristics of the used hierarchic data structure. The underlying principle will be based on a diffusion process, mimicking the simulated physics. The picture becomes even more involved when including adaptive time-stepping schemes, introducing grids with varying computational load impact, in addition to the varying impact from  message exchanges due to the hierarchy.

\section*{Acknowledgment}
The authors gratefully acknowledge the computing time granted by the Leibniz Supercomputing Centre (LRZ). Without the kind support, parts of this work would not have been possible.

\bibliographystyle{IEEEtran}
\bibliography{IEEEabrv,paper}

\begin{thebibliography}{1}
\providecommand{\url}[1]{#1}
\csname url@samestyle\endcsname
\providecommand{\newblock}{\relax}
\providecommand{\bibinfo}[2]{#2}
\providecommand{\BIBentrySTDinterwordspacing}{\spaceskip=0pt\relax}
\providecommand{\BIBentryALTinterwordstretchfactor}{4}
\providecommand{\BIBentryALTinterwordspacing}{\spaceskip=\fontdimen2\font plus
\BIBentryALTinterwordstretchfactor\fontdimen3\font minus
  \fontdimen4\font\relax}
\providecommand{\BIBforeignlanguage}[2]{{%
\expandafter\ifx\csname l@#1\endcsname\relax
\typeout{** WARNING: IEEEtran.bst: No hyphenation pattern has been}%
\typeout{** loaded for the language `#1'. Using the pattern for}%
\typeout{** the default language instead.}%
\else
\language=\csname l@#1\endcsname
\fi
#2}}
\providecommand{\BIBdecl}{\relax}
\BIBdecl

\bibitem{bader2012space}
M.~Bader, \emph{Space-filling curves: an introduction with applications in
  scientific computing}.\hskip 1em plus 0.5em minus 0.4em\relax Springer
  Science \& Business Media, 2012, vol.~9.

\bibitem{luitjens2007parallel}
J.~Luitjens, M.~Berzins, and T.~Henderson, ``Parallel space-filling curve
  generation through sorting,'' \emph{Concurrency and Computation: Practice and
  Experience}, vol.~19, no.~10, pp. 1387--1402, 2007.

\bibitem{bib:Frisch2014}
J.~Frisch, ``Towards massive parallel fluid flow simulations in computational
  engineering,'' Ph.D. dissertation, Technische Universit\"at M\"unchen, 2014.

\bibitem{bib:babufrmu2002}
M.~Bader, H.-J. Bungartz, A.~Frank, and R.-P. Mundani, ``Space tree structures
  for {PDE} software,'' in \emph{Computational Science}, ser. LNCS 2331,
  P.~Sloot, C.~Tan, J.~Dongarra, and A.~Hoekstra, Eds.\hskip 1em plus 0.5em
  minus 0.4em\relax Springer, 2002, pp. 662--671.

\bibitem{bib:brandt1977}
A.~Brandt, ``Multi-level adaptive solutions to boundary-value problems,''
  \emph{Mathematics of Computation}, vol.~31, no. 138, pp. 333--390, 1977.

\bibitem{bib:frmura2013}
J.~Frisch, R.-P. Mundani, and E.~Rank, ``Parallel multi-grid like solver for
  the pressure {P}oisson equation in fluid flow applications,'' in \emph{Proc.
  of the IADIS Int. Conf. on Applied Computing}.\hskip 1em plus 0.5em minus
  0.4em\relax IADIS Press, 2013, pp. 139--146.

\bibitem{bib:mufrvara2015}
R.-P. Mundani, J.~Frisch, V.~Varduhn, and E.~Rank, ``A sliding window technique
  for interactive high-performance computing scenarios,'' \emph{Advances in
  Engineering Software}, vol.~84, pp. 21--30, 2015.

\bibitem{bib:MPI}
\BIBentryALTinterwordspacing
{Message Passing Interface Forum}, ``Mpi: A message-passing interface standard,
  version 3.1,'' University of Tennessee, Knoxville, Tennessee, Specification,
  June 2015. [Online]. Available:
  \url{http://mpi-forum.org/docs/mpi-3.1/mpi31-report.pdf}
\BIBentrySTDinterwordspacing

\bibitem{bib:cybenko89}
G.~Cybenko, ``Dynamic load balancing for distributed memory multiprocessors,''
  \emph{Journal of parallel and distributed computing}, vol.~7, no.~2, pp.
  279--301, 1989.

\end{thebibliography}

\end{document}